\documentclass[5p,twocolumn,times,number]{elsarticle}

\usepackage{amsmath}
\usepackage{widetext}

\usepackage{graphicx,isolatin1}
\usepackage[squaren]{SIunits}

 \usepackage{etoolbox}

\newcommand{\textroman}[1]{\mbox{\rm #1}}
\newcommand{\gfrac}[2]{\displaystyle\frac{#1}{#2}}

\newcommand{\dd}{\mbox{d}}

 \newcommand{\mine}{\textroman{\scriptsize min}}
 \newcommand{\tot}{\textroman{\scriptsize tot}}
 \newcommand{\maxe}{\textroman{\scriptsize max}}

 \newcommand{\range}{\textroman{\scriptsize r}}

\RequirePackage{xspace}

\def\threshold{\text{threshold}}

\journal{Nucl.\ Instrum.\ Meth.\ A }

\begin{document}

\begin{frontmatter}

\title{A 5D, polarised, Bethe-Heitler event generator for $\gamma \to e^+e^-$ conversion}

\author[add1]{D.~Bernard\corref{cor}}
\ead{denis.bernard at in2p3.fr}

\address[add1]{LLR, Ecole Polytechnique, CNRS/IN2P3,
91128 Palaiseau, France}

\cortext[cor]{Tel 33 1 6933 5534}

\begin{abstract}
We describe a new version of the 5D, exact, polarised, Bethe-Heitler event
generator of $\gamma$-ray conversions to $e^+e^-$, developed
in the context of the HARPO project, that is able to simulate
successive events with different photon energies and on different atomic
targets without any substantial CPU overhead.
The strong correlation between kinematic variables in the divergence
of the five-dimensional differential cross section are mitigated by
performing each step of the conversion in the appropriate Lorentz
frame.
We extend the verification range down to 1\,keV above threshold and up to 1\,EeV.
This work could pave the way to the precise simulation of the
high-performance $\gamma$-ray telescopes and polarimeters of the
post-{\it Fermi}-LAT area.
\end{abstract}
 

\end{frontmatter}

\section{Introduction}

The Geant4 toolkit is used for the detailed simulation of many
scientific experiments that involve the interaction of elementary
particles with a detector 
\cite{Agostinelli:2002hh,Allison:2016lfl}.
For experiments with a simple geometry and that involve only
electrons, positrons and photons, the EGS5 framework is also
used \cite{Hirayama:2005zm,Bielajew}.
Radiation transport can also be simulated with the EGSnrc
\cite{EGSnrc}, Penelope \cite{Salvat:2015lni}, MCNP
\cite{Thompson:1979md} and FLUKA \cite{Battistoni} softwares.

The physics models of Geant4 and of EGS5 that describe the conversion
of a high-energy photon to an electron-positron pair, $\gamma \to
e^+e^-$, have been proven to be appropriate for the simulation of
electromagnetic (EM) showers
\cite{Nelson:2006te,Apostolakis:2015elm}.
These models are based on a number of approximations that limit their
domains of validity \cite{Gros:2016zst}:
\begin{itemize}
\item
They sample a product of independent one-dimensional (1D) probability density
functions (pdf), not the five-dimensional (5D) pdf.
\item They do not explicitly generate the target recoil momentum.
The leptons, that is, the electron and the positron, are therefore
generated incorrectly in the conversion plane.
 \item The polar angles of the electron and of the positron are
 generated independantly, so energy-momentum is not conserved.
This momentum imbalance artificially creates a transverse momentum
that the user could wrongly consider to be the recoil momentum, but
the momentum distribution so obtained is obviously completely
different from that predicted by QED.
 
\item In most models, high-energy and/or low-polar-angle
 approximations are used.
\end{itemize} 

Most pair telescopes, i.e., those detectors that detect the conversion of
high-energy photons to a pair, consist of an active target, i.e., a
detector in which the photon converts and the trajectories of the
electron and of the positron (``tracks'') are measured, eventually
followed by a calorimeter, in which the energy of the two leptons is
measured.
The photon angular resolution of pair telescopes is the
combination of several contributions, that include 
\begin{itemize}
\item the single-track angular resolution induced by the multiple
 scattering undergone by the lepton in the detector;
\item the incorrect calculation of the momentum of the incident photon
 as the sum of the momenta of the final state particles, because the
 energy of the recoiling nucleus is too small to produce a measurable
 track in the detector, and the contribution of its momentum to the sum is
 then missing.
\end{itemize} 

The existing Geant4 and EGS5 physics models are appropriate for the
description of the conversion of photons in the past and present
$\gamma$-ray telescopes, for which the multiple-scattering
contribution dominates: the inaccurate simulation of the
kinematics of the conversion and the
differences between physics models are washed out by multiple
scattering (Fig. 8 bottom of \cite{Gros:2016zst}).

Since the multiple scattering RMS of the angle of a charged particle
traversing a slab of matter varies as the inverse
of the track momentum, the photon angular resolution of pair
telescopes degrades badly at low photon energy, that is, at low track
momentum \cite{Ackermann:2012kna}.
When an optimal Kalman-filter-based tracking is applied to the tracks,
the multiple scattering contribution to the 
 photon angular resolution is found to vary as $E^{-3/4}$
\cite{Bernard:2012uf}.
A coarser angular resolution affects the effectiveness of background
rejection in the selection of the photons associated with a given
source, and therefore degrades the point-like source sensitivity of
the instrument.

A number of techniques are being developed to improve on the angular
resolution with respect to that of the {\it Fermi}-LAT
\cite{Ackermann:2012kna}. A factor of three at 100\,MeV
can be obtained by using
all-silicon active targets, that is, without high-$Z$
converters \cite{DeAngelis:2016slk,AMEGO} or with emulsions
\cite{Takahashi:2015jza,Ozaki:2016gvw}.
With even lower-density homogeneous detectors such as gas
time-projection chambers (TPC) a factor of ten improvement can be
achieved \cite{Bernard:2012uf} and the single-track angular resolution is so good that
polarimetry has been predicted to be possible despite the dilution of
the polarisation asymmetry induced by multiple scattering
\cite{Bernard:2013jea}, and has actually been demonstrated by the
characterisation of a TPC prototype in a particle beam
\cite{Gros:2017wyj}.

For such high-performance telescopes, the single-track angular
resolution is so good that the inaccuracy of the event generator
becomes the main bias in the understanding of the photon
angular resolution
(Fig. 8 up-right of \cite{Gros:2016zst}).
Furthermore, attempts to verify the generation of the conversion of
linearly polarised photons by the physics model of Geant4 have shown
a strong departure with respect to the QED prediction
 (Fig. 11 of \cite{Gros:2016zst}).

In the context of the HARPO project, we have written an exact, 5D,
polarised Bethe-Heitler event generator \cite{Bernard:2013jea}.
The main difficulty in sampling the 5D differential cross section
is that it diverges at small $q^2$
(where $q$ is the momentum ``transferred'' to the target) 
and, for high energy, at small
lepton polar angles.
We first \cite{Bernard:2013jea} solved this issue
by using the BASES/SPRING 
\cite{Kawabata:1995th} 
implementation of the VEGAS method 
\cite{Lepage:1977sw}.
This software proved to be excellent for the performance studies of
the HARPO project, for which samples of many events were
generated and then studied for given photon energy and
target nucleus.
But the VEGAS method first optimizes a segmentation of each of the
five variable ranges corresponding to the 5D phase-space that
defines the final state, after which the differential cross section is
tabulated on the 5D grid obtained.
These calculations prior to the generation of the first event need
several seconds of CPU time which is obviously not acceptable for a particle
physics simulation package like Geant4 for which the user needs to
generate sequentially events with various photon energies and on
various targets.
The photon can either interact
\begin{itemize}
\item with a nucleus (nuclear conversion)

 $\gamma ~ Z ~ \to ~ e^{+}e^{-} ~ Z $;

\item or with an electron (triplet conversion) 

 $\gamma ~ e^{-} ~ \to ~ e^{+}e^{-} ~ e^{-} $.
\end{itemize}

This paper describes a new event generator that does not use the VEGAS
method. However, it uses the same specification list as that described
in \cite{Bernard:2013jea} for the VEGAS-based generator (5D sampling,
polarised, nuclear or triplet, no approximation, energy momentum
conservation).
After some general considerations on $\gamma$-ray conversions to pairs
and their event generation (Sect. \ref{sec:gen}), we describe the part of
the structure that is common to the two codes
(Sect. \ref{sec:common}).
In section \ref{sec:VEGAS}, we present some properties of the
VEGAS-based generator, and in section \ref{sec:one-shot}, those of the
new version.

\section{Bethe-Heitler, 5D, polarised, event generation}
\label{sec:gen}

The non-polarised differential cross section was obtained by
Bethe \& Heitler \cite{Bethe-Heitler}, after which the calculation of
the polarised cross section opened the way to polarimetry with pair
conversions \cite{BerlinMadansky1950,May1951,jau}.
The final state can be defined by the polar angles $\theta_+$ and
$\theta_-$, and the azimuthal angles $\phi_+$ and $\phi_-$, of the
electron (-) and of the positron (+), respectively, and the fraction
$x_+$ of the energy of the incident photon carried away by the
positron, $x_+ \equiv E_+/E$
(Table\,\ref{tab:BH:var}.
A schema can be found in Fig. 3 in \cite{Bernard:2013jea}).
All these expressions are referred to as "Bethe-Heitler" in this paper.

\begin{table} \small \caption{Bethe-Heitler variable list, all in the laboratory frame.\label{tab:BH:var} }
\begin{tabular}{llllll} \hline \hline 
1 & $\theta_+$ & positron polar angle\\
2 & $\theta_-$ & electron polar angle\\
3 & $\phi_+$ & positron azimuthal angle\\
4 & $\phi_-$ & electron azimuthal angle\\
5 & $x_+$ & fraction of the photon energy carried by the positron\\ \hline \hline 
\end{tabular}
\end{table}

The two dominant Feynman diagrams were taken into account in \cite{Bethe-Heitler}, which
is an excellent approximation for nuclear conversion and for
high-energy triplet conversion \footnote{These two dominant digrams
 are often named the Borsellino diagrams, see the dicussion of Fig. 1
 in \cite{Mork1967}.}:
for photon energies smaller than $10\,m c^2$, where $m$ is the
electron mass, neglecting the $\gamma-e$ exchange diagrams induces a
relative change in the triplet total cross section smaller than 7\,\%
\cite{Mork1967}.
Only the linear polarisation of the incoming photon takes part in
these expressions.
The circular polarisation of the incoming photon does not take part at
this first order of the Born approximation and therefore no polarisation is
transferred to the leptons.

In the case where the conversion takes place in the field of an
isolated, ``raw'', nucleus or electron, the bare Bethe-Heitler
expression mentioned above is used.
In the case where the nucleus or the electron are part of an atom, the
screening of the target field by the other electrons of the atom is
described by a simple form factor, function of $q^2$ \cite{Mott:1934}
(nuclear) or \cite{WheelerLamb1939} (triplet).

This work was performed under a series of approximations: 
\begin{itemize}
\item 
The nuclear form factor that affects the probability of very large-$q^2$
events is not considered.
\item Landau-Pomeranchuk-Migdal (LPM)
\cite{Landau:1953um,Migdal:1956tc} suppression effects in the
 differential cross-section at very high-energy are not
 considered.
\item 
Any pre-existing non-zero momentum of the target prior to the
conversion, such as in the case of Compton ``Doppler'' broadening, is
not considered either.
\item 
Coulomb corrections, due to the electromagnetic interaction between
the final-state charged particles, that is prevalent, in particular,
close to threshold, is not addressed.
\end{itemize}

For triplet conversion, the experimental issue of
deciding which of the two negative electrons in the final state
belongs to the pair and which one is ``recoiling'' is irrelevant here:
by construction, 
the variables $\phi_-$ and $\theta_-$ that take part in the
Bethe-Heitler expressions refer to the electron of the pair, while
$q^2$ refers to the recoil.

\section{The two event generators: common properties}
\label{sec:common}

The second difficulty originates from the fact that the kinematic
variables from which the differential cross section is computed cannot
be computed from $q^2$, the value of which drives the main divergence:
the kinematic variables must be generated at random first, and then
$q^2$ be computed from them, so some of the five variables turn out to
be strongly correlated with each other.
This is a general issue, even when the VEGAS method is used, as it is
based on the assumption that the pdf is not too different from a product of
1D pdfs, so that a simple product of 1D segmentations of the phase
space can be used.

The situation improves when computations are performed in the
appropriate Lorentz systems as follows: 

\begin{table} \small
\caption{Kinematic variables and the Lorentz frame in which they are defined.
\label{tab:var} }
\begin{tabular}{llllll} \hline \hline 
1 & $\theta$ & target and pair polar angle & CMS\\
2 & $\mu$ & $e^+ e^-$ invariant mass & \\
3 & $\theta_\ell$ & $e^+$ and $e^-$ polar angle & pair frame\\
4 & $\phi_\ell$ & $e^+$ and $e^-$ azimuthal angle & pair frame\\
5 & $\phi$ & target and pair azimuthal angle & CMS \\ \hline \hline 
\end{tabular}
\end{table}

\begin{itemize}
\item The center-of-mass system (CMS) boost is determined from the
 values of the photon energy, $E$, and of the target mass,
 $M$.
\item The five kinematic variables are taken at random, namely 
 $\theta$, 
 $\mu$, 
 $\theta_\ell$,
 $\phi_\ell$ and
 $\phi$, as defined in Table \ref{tab:var}.

\item 
In the CMS, the target (with mass $M$) and the pair (with mass $\mu$)
have opposite momenta. Their 4-vectors are computed.

\item
The ``decay'' of the pair to an electron and a positron is
performed in the pair center-of-mass frame.
Their 4-vectors are computed from the pair invariant mass, $\mu$, and
from the ``decay'' polar and azimuthal angles, $\theta_\ell$ and
$\phi_\ell$. $\ell$ stands for lepton.

\item The lepton 4-vectors are boosted ``back'' to the CMS.

\item The three final particle 4-vectors
(electron, positron, target) are boosted ``back'' to the laboratory
 frame.

\item The Bethe-Heitler variables, including the polar and azimuthal
 angles of each particle, are obtained from the 4-vectors.

\item Then the pdf for that event candidate is computed.
\end{itemize}

As the $z$ axis is defined by the direction of the incident photon,
the target recoil momentum and the pair momentum are defining a plane
that contains the photon momentum: their azimuthal angle are left
unchanged upon the boost and are equal to $\pm \phi$.

Given the chosen set of kinematic variables, the final-state phase
space is normalised as described in eq. (47.20) of Sect. 47.4.3 of
\cite{Patrignani:2016xqp}), and the differential cross section becomes 
\begin{eqnarray}
 \dd \sigma & = &
 \gfrac{1}{(2\pi)^5} \gfrac{1}{32 \, M\sqrt{s} \, E} \left|{\cal M}\right|^2
 \left| p_+^\ast \right|
 \,
 \left| p_r \right|
 \,
 \dd \mu
 \,
 \dd \Omega_+^\ast 
 \,
 \dd \Omega_r,
\end{eqnarray}
where
$( p_+^\ast , \Omega_+^\ast )$ refers to the kinematic variables of the
positron in the pair rest frame and
$( p_r, \Omega_r )$ to the kinematic variables of the target recoil in the CMS.
We obtain:
\begin{eqnarray}
\dd \sigma = H \, (X_{u} + P \, X_{p})
 \,
\dd \mu
 \,
 \dd \Omega_+^\ast 
 \,
 \dd \Omega_r,
\label{eq:dsig:short}
\end{eqnarray}
 with:
\begin{eqnarray}
 H = \frac{- \alpha \, Z^2 r_0^2 }{ (2 \pi)^2 }
 \frac{ \left| p_+^\ast \right| \left| p_r \right| \, m^2 M}{E^3 \sqrt{s}\left|\vec{q}\right|^4} ,
\label{eq:pair:BigPhi:b}
\end{eqnarray}

 \clearpage
\begin{widetext}
\begin{eqnarray}
X_{u}& = &
\left[ 
\left(
\frac{p_+ \sin{\theta_+}}{E_+ - p_+ \cos{\theta_+}}
\right)^2 (4 E_-^2 - q^2)
+
\left(
\frac{p_- \sin{\theta_-}}{E_- - p_- \cos{\theta_-}}
\right)^2 (4 E_+^2 - q^2)
+
\right.
\nonumber 
\\
 & & 
 \left. 
\frac{2p_+ p_- \sin{\theta_+} \sin{\theta_-} \cos{(\varphi_+-\varphi_-)}}{(E_- - p_- \cos{\theta_-})(E_+ - p_+ \cos{\theta_+})}
(4 E_+ E_- + q^2 -2 E ^2) 
- 
2 E ^2
\frac{(p_+ \sin{\theta_+})^2 + (p_- \sin{\theta_-})^2}
{(E_+ - p_+ \cos{\theta_+})(E_- - p_- \cos{\theta_-})}
\right] ,
\label{eq:pair:unpol:X}
\end{eqnarray}
\begin{eqnarray}
X_{p} & = & 
\cos{2\varphi_-} 
(4 E_+^2 -q^2)
 \left(
\frac{p_- \sin{\theta_-}}{E_- - p_- \cos{\theta_-}} \right)^2
+
\cos{2\varphi_+} 
(4 E_-^2 -q^2)
 \left(
 \frac{p_+ \sin{\theta_+}}{E_+ - p_+ \cos{\theta_+}}
\right)^2 
 \nonumber 
\\
 & & 
+2 \cos{(\varphi_++\varphi_-)} (4 E_+ E_- +q^2)
\frac{p_- \sin{\theta_-} p_+ \sin{\theta_+}}{(E_- - p_- \cos{\theta_-})(E_+ - p_+ \cos{\theta_+})} .
\label{eq:pair:pol:X}
\end{eqnarray}
 \end{widetext}
 
A number of technical verifications were undertaken to check that the
whole process was implemented correctly, including a comparison of the
result of a calculation of the pair invariant mass based on the
Bethe-Heitler variables to the value of $ \mu$.

\section{The VEGAS-based generator}
\label{sec:VEGAS}

VEGAS is a multipurpose algorithm for multidimensional
integration that was developed for particle-physics applications.
It is adaptative in that it automatically concentrates evaluations of
the integrand in those regions where it is largest in
magnitude \cite{Lepage:1977sw,Lepage:1980dq}.
The integral is performed on an $n$D grid that is the simple product of
$n$ 1D grids (here $n=5$). Initially, the segmentation is uniform on
each 1D axis.
It is then iteratively modified so as to minimize the variance of the
estimation of the integral.

We have used the VEGAS method in its BASES/SPRING implementation 
\cite{Kawabata:1995th} to build the first generator
\cite{Bernard:2013jea}.
After the grid optimisation has been performed by BASES, the
pdf, i.e., the normalised differential
cross section, is known on all nodes of the $n$D grid.
The SPRING package then generates the exact pdf from the tabulated one
by the acceptance-rejection method (see, example, Sect. 40.3 of
\cite{Patrignani:2016xqp}).

If the pdf has singularities that involve several of the variables on
which the generation is performed in a correlated way, the grid
optimisation does not converge to an appropriate solution and the
VEGAS method is inefficient \cite{Kawabata:1995th}.
The choice of variables documented in the previous section had been
made with that limitation in mind.
Some of the five variables that were actually used in
\cite{Bernard:2013jea} were functions of the variables listed in Table
\ref{tab:var}, so as to minimize any 1D pdf divergence for that variable.
This was done for practical purpose only (plotting etc.) since the VEGAS
method deals with divergences effectively, within the correlation
limitation already mentioned.
This change of variables is, therefore, not detailed here.

A different set of variables is used for the new generator for which
the resampling facility of the VEGAS method is not available anymore,
as presented in the next section.

\section{The new event generator}
\label{sec:one-shot}

From the study of the 1D distributions of the five variables provided
by the VEGAS-based generator, we obtained a set of variables $x_i,
i=1\cdots 5$ so that each $x_i$ takes values on a (finite length)
segment and its pdf does not have any singularity.

\begin{figure}[ht] 
\includegraphics[width=\linewidth]{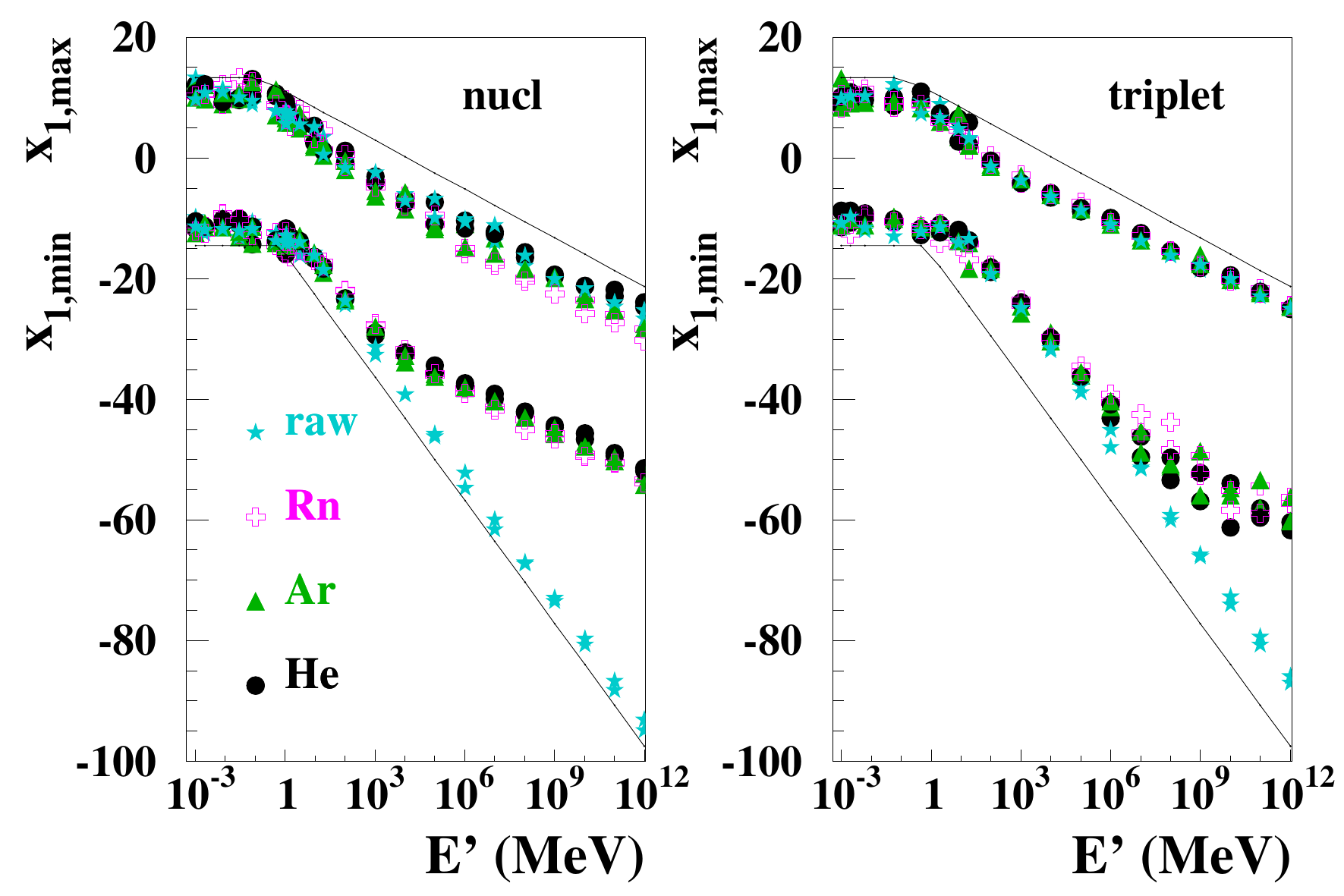}
\caption{Variation with $E'$ of the maximum and of the minimum values
 for variable $x_1$.
Left: nuclear conversion; Right: triplet conversion.
``raw'' isolated charged target (star), and the following atoms:
helium (bullet),
argon (upper triangle),
radon (plusses).
The continuous lines denote the bounds that are used by the generator.
$P=1$ and $P=0$ samples are plotted for each photon-energy value.
\label{fig:bounds}
 }
\end{figure}

\begin{figure*}[ht] 
\includegraphics[width=0.47\linewidth]{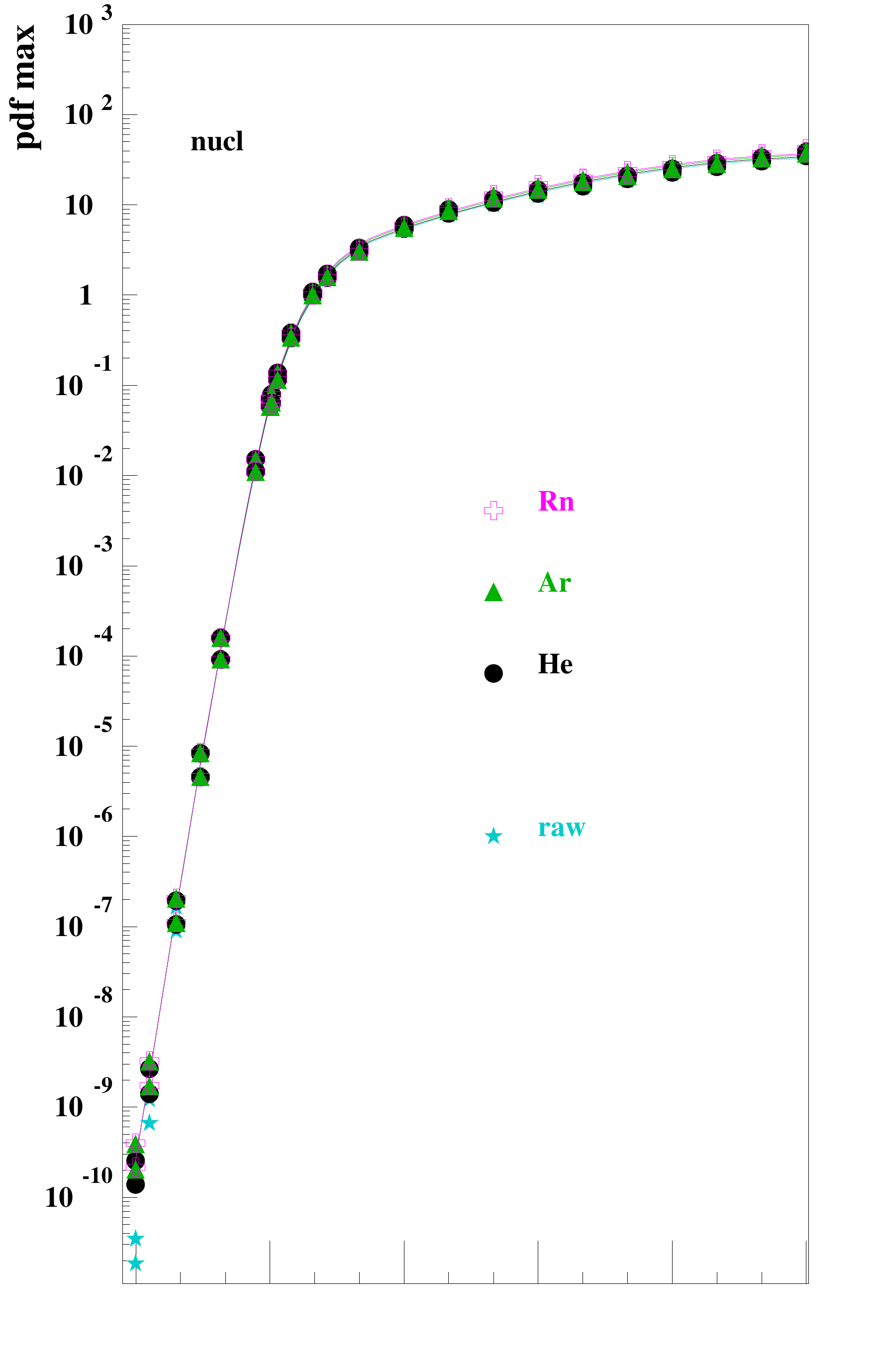}
\includegraphics[width=0.47\linewidth]{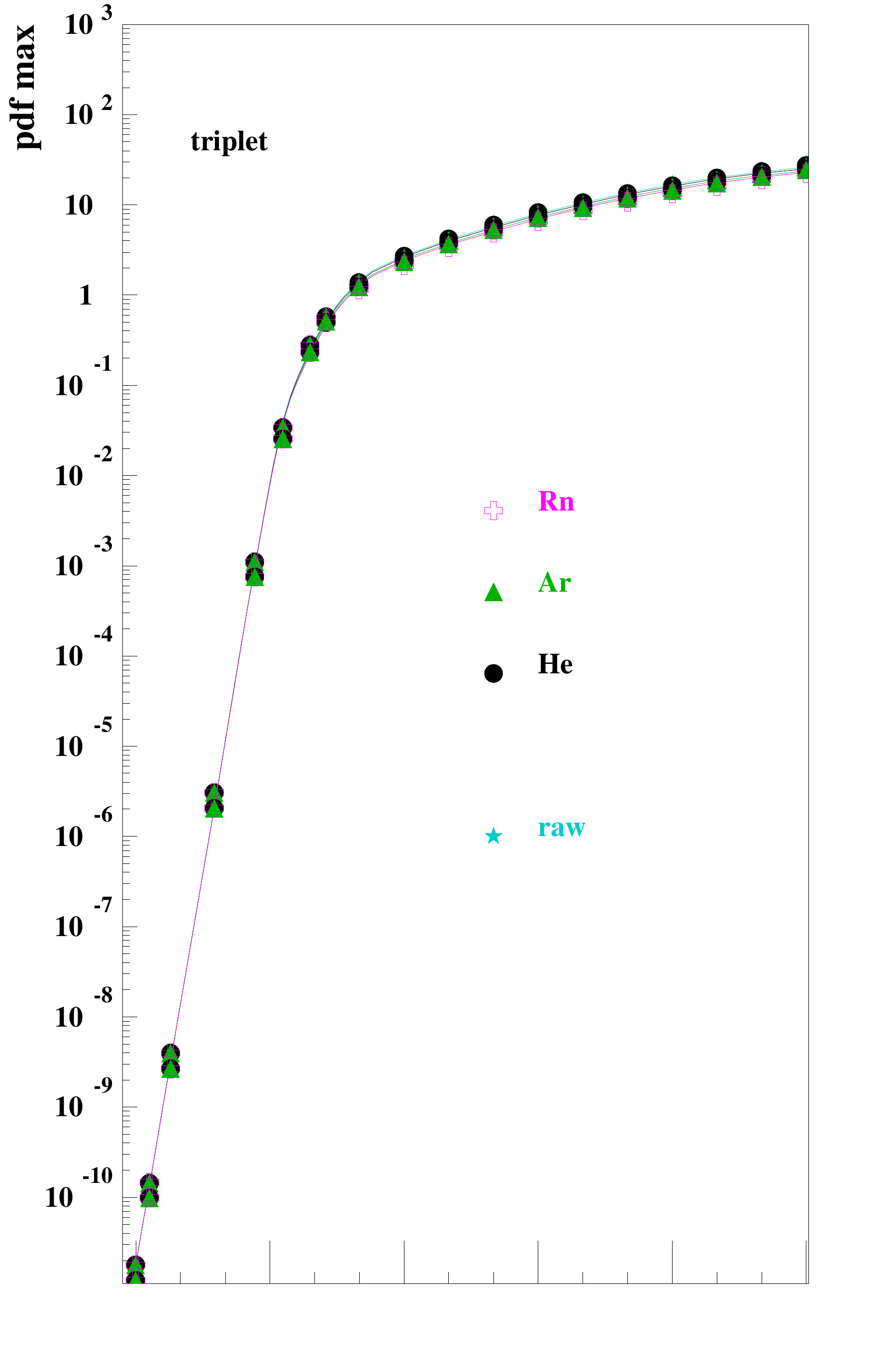}

\vspace{-1.14cm}
\includegraphics[width=0.47\linewidth]{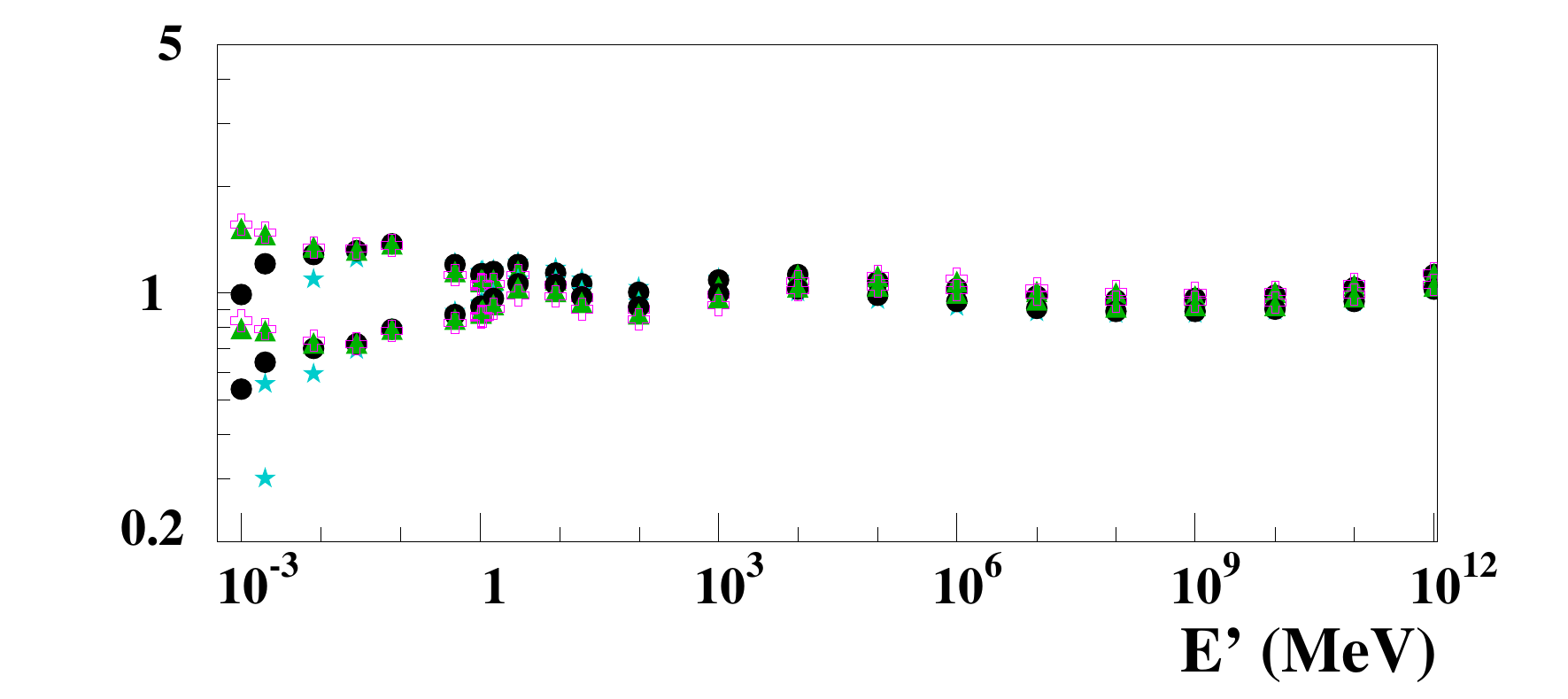}
\includegraphics[width=0.47\linewidth]{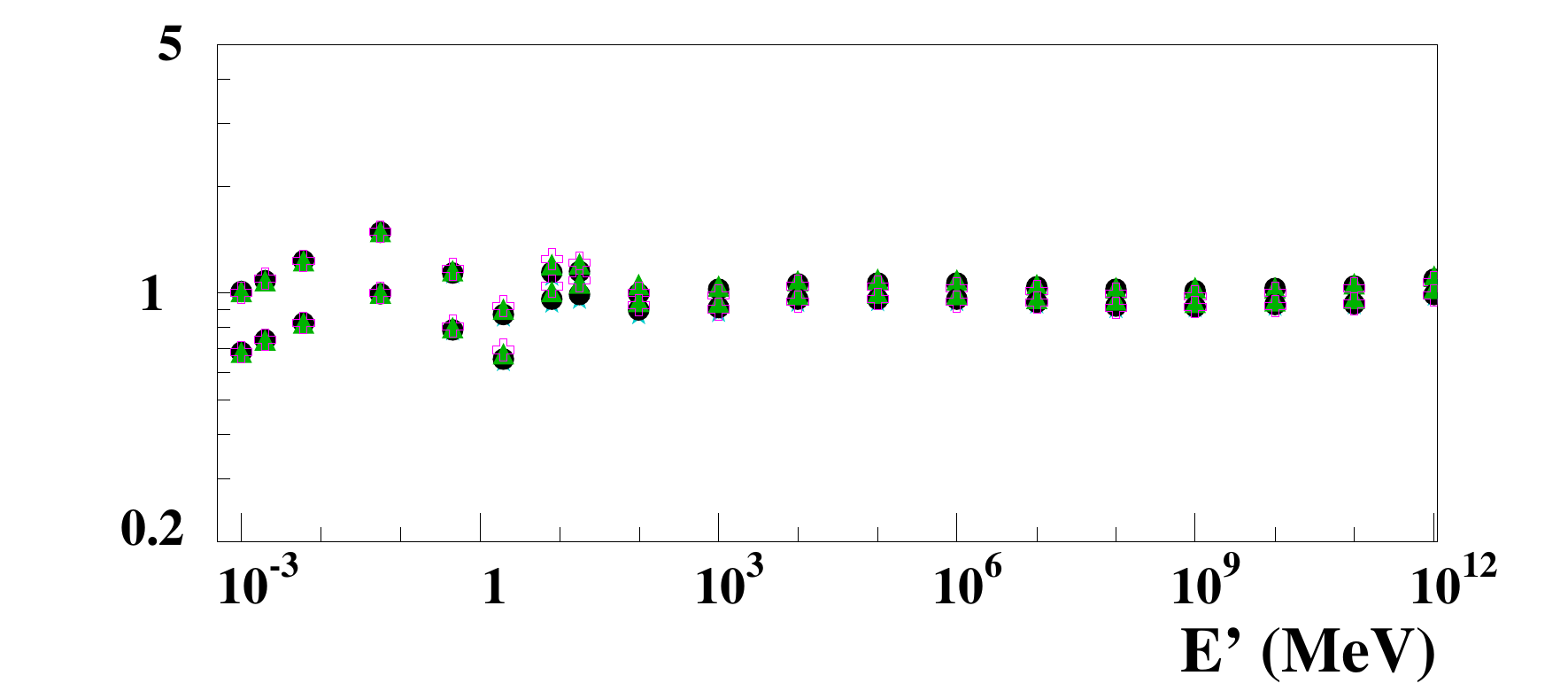}
\caption{Top: Variation of the maximum value of the pdf
 (arbitrary units) with $E'$.
Left: nuclear conversion; Right: triplet conversion.
``raw'' isolated charged target (star), and the following atoms:
helium (bullet),
argon (upper triangle),
radon (plusses).
The continuous lines denote the parametrisation of eq. (\ref{eq:fit:diff:max}).
Bottom: residues of the fit of eq. (\ref{eq:fit:diff:max}).
$P=1$ and $P=0$ samples are plotted for each photon-energy value.
\label{fig:max:diffcross}
 }
\end{figure*}

\begin{figure}[ht] 
\includegraphics[width=\linewidth]{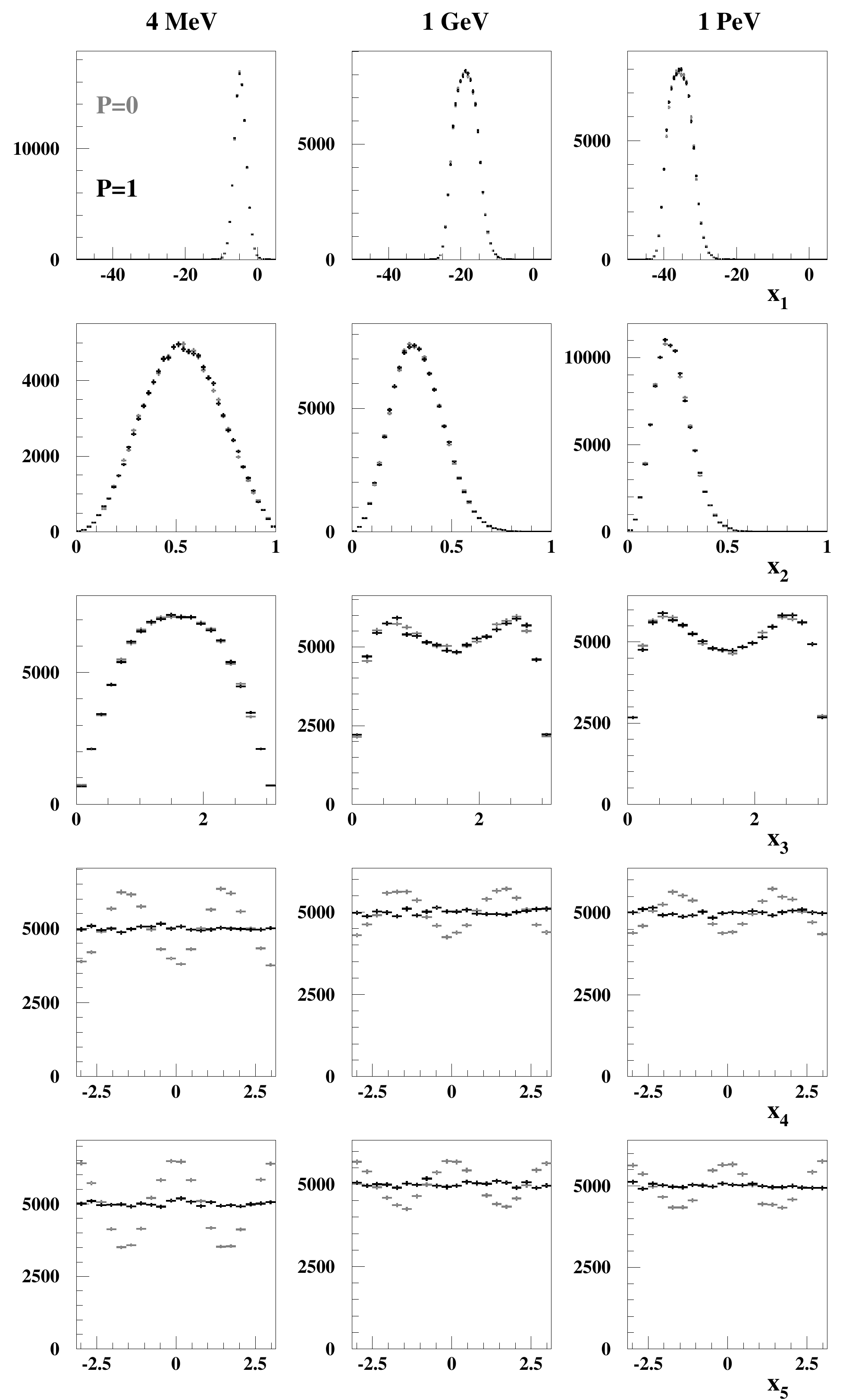}
\caption{
The distributions of the five generation variables for the nuclear conversion of $\gamma$-rays of energies
$4\,\mega\electronvolt$, 
$1\,\giga\electronvolt$ and
$1\,\peta\electronvolt$
on argon.
$P=1$ (grey) and $P=0$ (black) samples are plotted for each photon-energy value.
\label{fig:argon}
 }
\end{figure}
\begin{table} 
\begin{center}
\small
\caption{Relationship between the generator variables, $x_i,
i=1\cdots 5$, and the kinematic variables, and their range.\label{tab:var:change} }
\begin{tabular}{llllll} \hline \hline
$i$ & & Jacobian & $x_i$ range \\ \hline \noalign{\vskip3pt}
1 & $\cos\theta = \gfrac{y - 1}{1 + y}$, $y = \exp(x_1)$ & $ \gfrac{y} {(1+y)^2}$ & $[x_{1l},x_{1u}]$ \\
2 & $\mu = \mu_{\mine} \, (\mu_{\range})^{x_2^2} $ & $2 \, x_2 \log{( \mu_{\range})} \mu$ & $[0,1]$ \\
3 & $\cos\theta_\ell = x_3$ & $|\sin \theta_\ell|$ & $[0,\pi]$ \\
4 & $\phi_\ell = x_4 $ & 1 & $[-\pi,\pi]$ \\
5 & $\phi = x_5 $ & 1 & $[-\pi,\pi]$ \\ \hline \hline 
\end{tabular}
\end{center}
\end{table}

The relation between the generator variables
$x_i, i=1\cdots 5$ and the kinematic variables is given in Table 
\ref{tab:var:change}.
$\mu_{\range} = \mu_{\maxe} / \mu_{\mine}$, with
$\mu_{\mine} = 2 \, m$ and
$\mu_{\maxe} = \sqrt{s} - M$.
In this document $\log$ refers to the natural logarithm.

For better readability, energy-variation plots are presented as a
function of
\begin{equation}
 E' = E - E_{\threshold},
\end{equation}
where the energy threshold is
\begin{itemize}
\item 
 $ E_{\threshold} = 2 m c^2$ for nuclear conversion and
\item 
 $E_{\threshold} = 4 m c^2$ for triplet conversion.
\end{itemize}
$E'$ is the available kinetic energy of the leptons for nuclear
conversion, but not for triplet conversion.

The $x_1$ bounds, $x_{1l},x_{1u}$, are given in Fig. \ref{fig:bounds}
as a function of $E'$.
For each energy, for each target, for each process (nuclear, triplet),
for each side (upper bound, lower bound), the two points correspond to
fully polarised ($P=1$) and to non-polarised ($P=0$) simulated data.
Each value is computed from a sample of $10^5$ simulated events.

We take $x$ to be the set of variables for a conversion event,
$x = (x_i,i=1\cdots 5)$, and $X$ the physically accessible part of the
5D space for $x$.
The variables, $x$, are taken at random with a
uniform pdf $p_0(x) = p_0$.
The exact pdf, $p(x)$, is then ``carved'' into that 5D-flat sample
using the acceptance-rejection method: if a constant $C$ can be found,
so that for all points $x$,
\begin{equation}
 C p_0(x) > p(x),
\end{equation}
then taking a random
variable $u$, flat in $[0,1]$, and accepting events for which
\begin{equation}
u C p_0(x) < p(x),
\end{equation}
provides an $x$ sample with pdf $p(x)$
(section 40.3 of \cite{Patrignani:2016xqp}).

For $p_0=1$, $C$ is the maximum value of $p(x)$ that we have studied
with the VEGAS-based generator.
It can be represented by the following function
\begin{equation}
\gfrac{\tau_1 E'^{\tau_3 + \tau_5 \log{E'}} }{\tau_2 E'^{\tau_6} + E'^{\tau_4}}
 \left(1 + \tau_8 Z^{\tau_9} \gfrac{Q}{1+Q}\right)
 \quad\mathrm{with}\quad
 Q = \gfrac{E'}{\tau_{10}},
 \label{eq:fit:diff:max}
\end{equation}
where the sets of constants $\tau_i$ are different for nuclear and for
triplet conversions.
These expressions of the values of $C(E',Z)$ for nuclear and triplet
conversions are compared to the values obtained from simulated samples
in Fig. \ref{fig:max:diffcross}.
In practice, for safety, $C$ is enlarged by a factor of 1.5 with
respect to what can be seen on Fig. \ref{fig:max:diffcross}.
In these expressions, $Z$ is the atomic number of the atom on which
the conversion took place.
The raw data, that is, the conversion on an isolated electron or
nucleus, are found to be well represented by an effective $Z = 0.5$.
The distributions of the five variables
do not show any remaining divergence (Fig. \ref{fig:argon}).

\subsection{Total cross section}

\begin{figure}[ht] 
\includegraphics[width=0.49\linewidth]{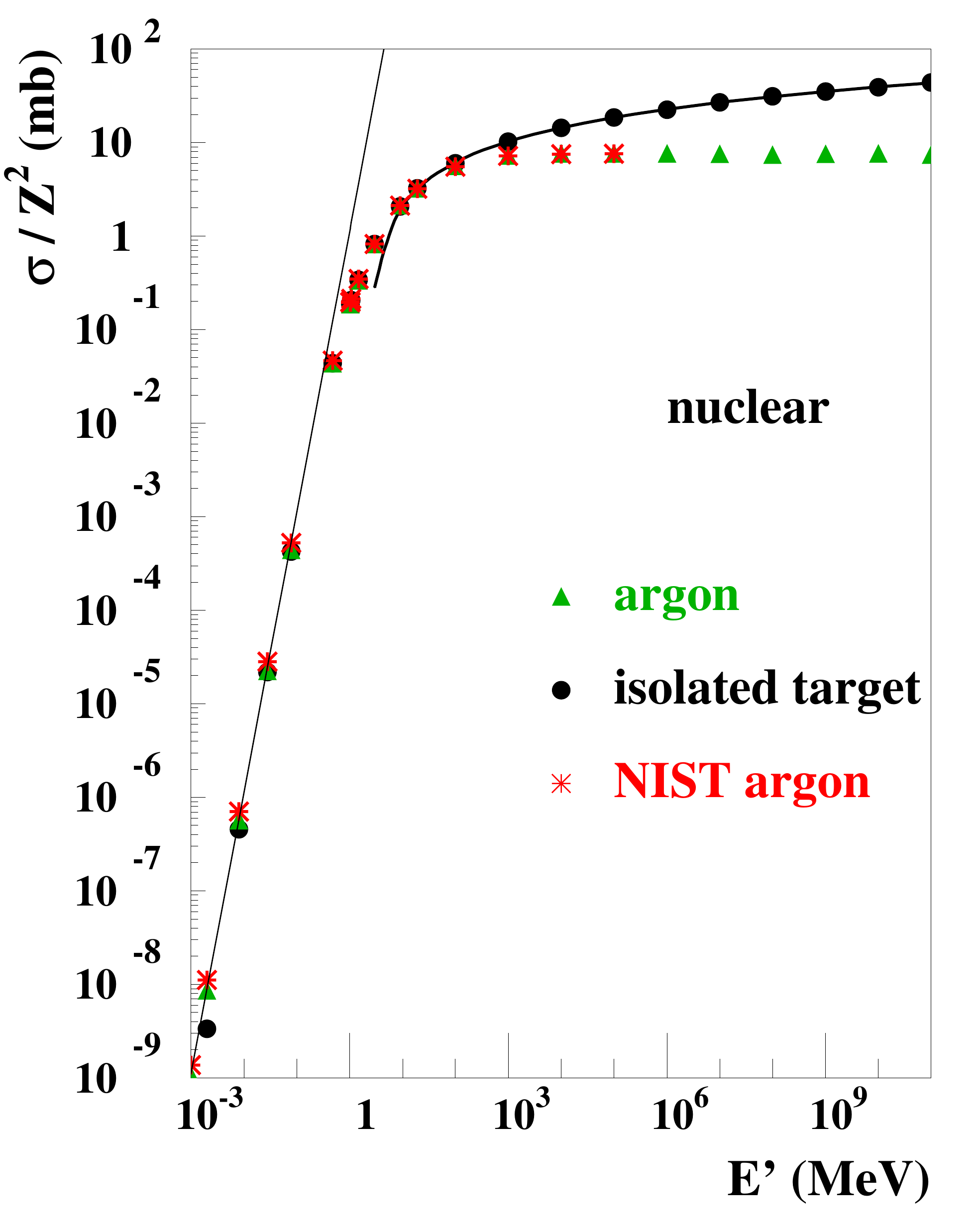}
\includegraphics[width=0.49\linewidth]{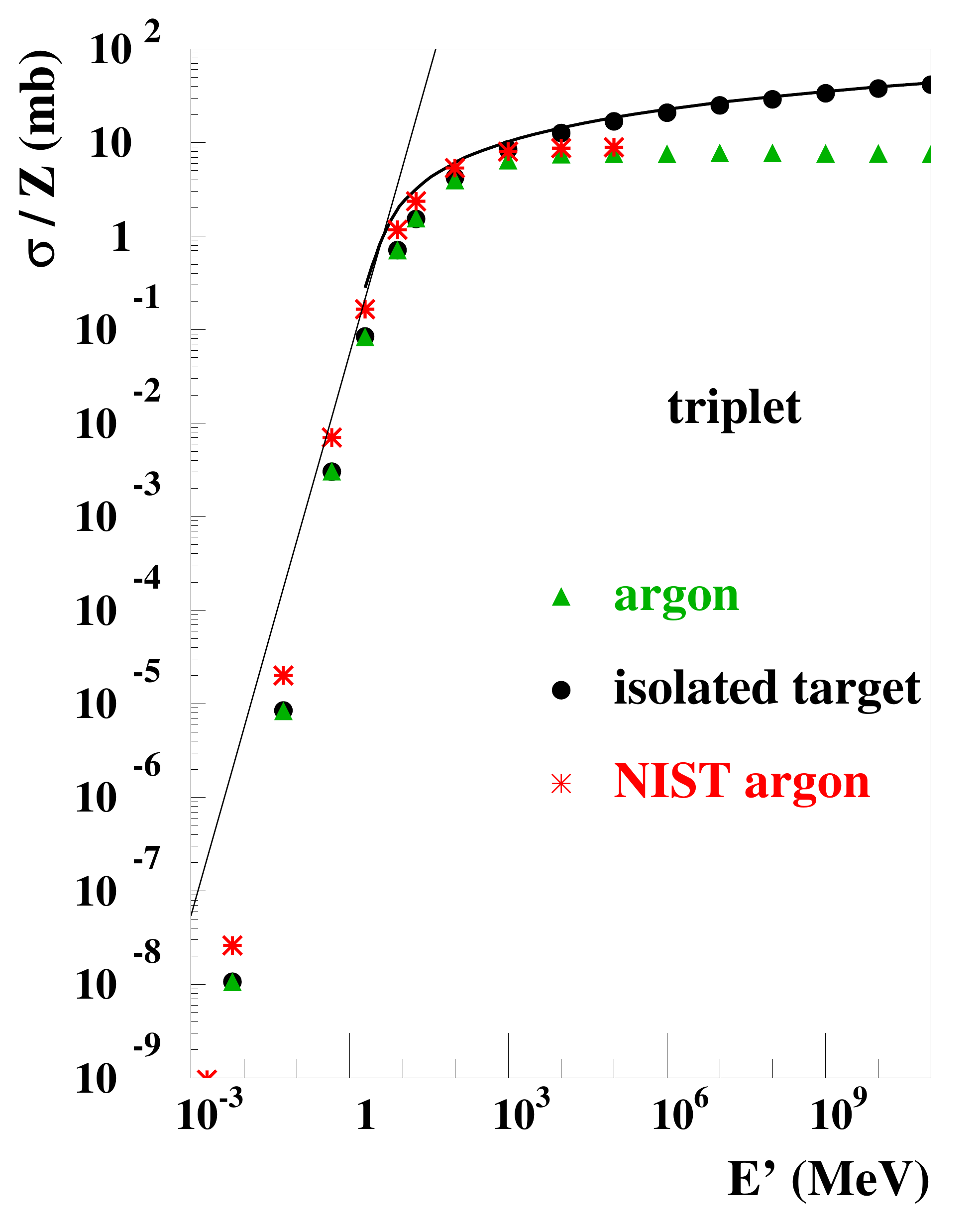}
\caption{Total cross section normalised to $Z=1$, as a function of $E'$.
Left: nuclear conversion.
Right: triplet conversion.
Thick line: high-energy approximation.
Thin line: low-energy approximation.
Integration of the differential cross section used in the generator for conversion on isolated charged particles 
(bullets) or argon atoms (upper triangles).
Values tabulated by NIST (crosses).
\label{fig:tot}
 }
\end{figure}

We integrate the differential cross section used by the generator and
obtain the total cross section (Fig. \ref{fig:tot}).

The thick line shows the high-energy approximation
\cite{Bethe-Heitler}
\begin{eqnarray}
\sigma_{\tot} =
\gfrac{28}{9}\log{\left(\gfrac{2 E}{m c^2}\right)} - \gfrac{218}{27}.
\label{eq:sig:tot:HE}
\end{eqnarray}

The thin line shows low-energy approximation.
For nuclear conversion 
\cite{Racah1934}
\begin{equation}
 \sigma_{\tot} = \alpha r_0^2 Z^2 \gfrac{\pi }{12} \left( \gfrac{E}{m c^2} -2 \right)^3
\end{equation}

For triplet conversion we use the expression obtained using the 
Borsellino diagrams \cite{Borsellino1947}, something that is appropriate for
this comparison to the Bethe-Heitler cross section based on the same
assumption:
\begin{equation}
 \sigma_{\tot} = \alpha r_0^2 \gfrac{\pi \sqrt{3}}{2^3 3^3}
 \left( \gfrac{E}{m c^2} -4 \right)^2.
\end{equation}
The misprint in \cite{Borsellino1947} mentioned in 
\cite{Votruba1948a} has been corrected.

The total cross section for isolated charged particles (bullets) can
be compared to the approximations (thin and thick lines).
The total cross section for conversion on a charged target inside an
argon atom (upper triangle) can be compared to the computations of the
National Institute of Standards and Technology (NIST) Physical
Reference Data and based on Ref. \cite{Hubbell1980}.

\begin{itemize}
\item At energies larger than $\approx 100\,\mega\electronvolt$, the
 raw cross section is found to be compatible with the high-energy
 approximation for both triplet and nuclear conversions.
\item Cross sections on atoms are found to be compatible with that
 tabulated by NIST except for very-low-energy triplet conversion for
 which a factor of $\approx 2.44$ is missing which is not surprising
 as the lack of the exchange diagrams in the Bethe-Heitler differential cross
 section is sensitive there (See the discussion in section 2 of
 Ref. \cite{JosephRohrlich1958}).
 
\item The low-energy nuclear cross section is found to be compatible
 with the low energy approximation, but the low-energy triplet cross
 section is not, both for the generator and for NIST data: these data
 seem to support an $ \left( E / m c^2 -4 \right)^3$
 dependence rather than the $ \left( E / m c^2 -4 \right)^2$
 dependence predicted by the Borsellino \cite{Borsellino1947} and by
 the Votruba \cite{Votruba1948a} expressions, something that is not
 understood.
\end{itemize}

\subsection{Distributions of kinematic variables}

We can now examine the distributions of variables that are not easily
accessible by the physics models that do not sample the full 5D
differential cross section.
We extend the verifications that were performed in the past with the
VEGAS-based generator \cite{Bernard:2013jea,Gros:2016zst,Gros:2016dmp}
to the larger energy range explored in this work with the new
generator.
\begin{figure}[ht] 
 \includegraphics[width=0.49\linewidth]{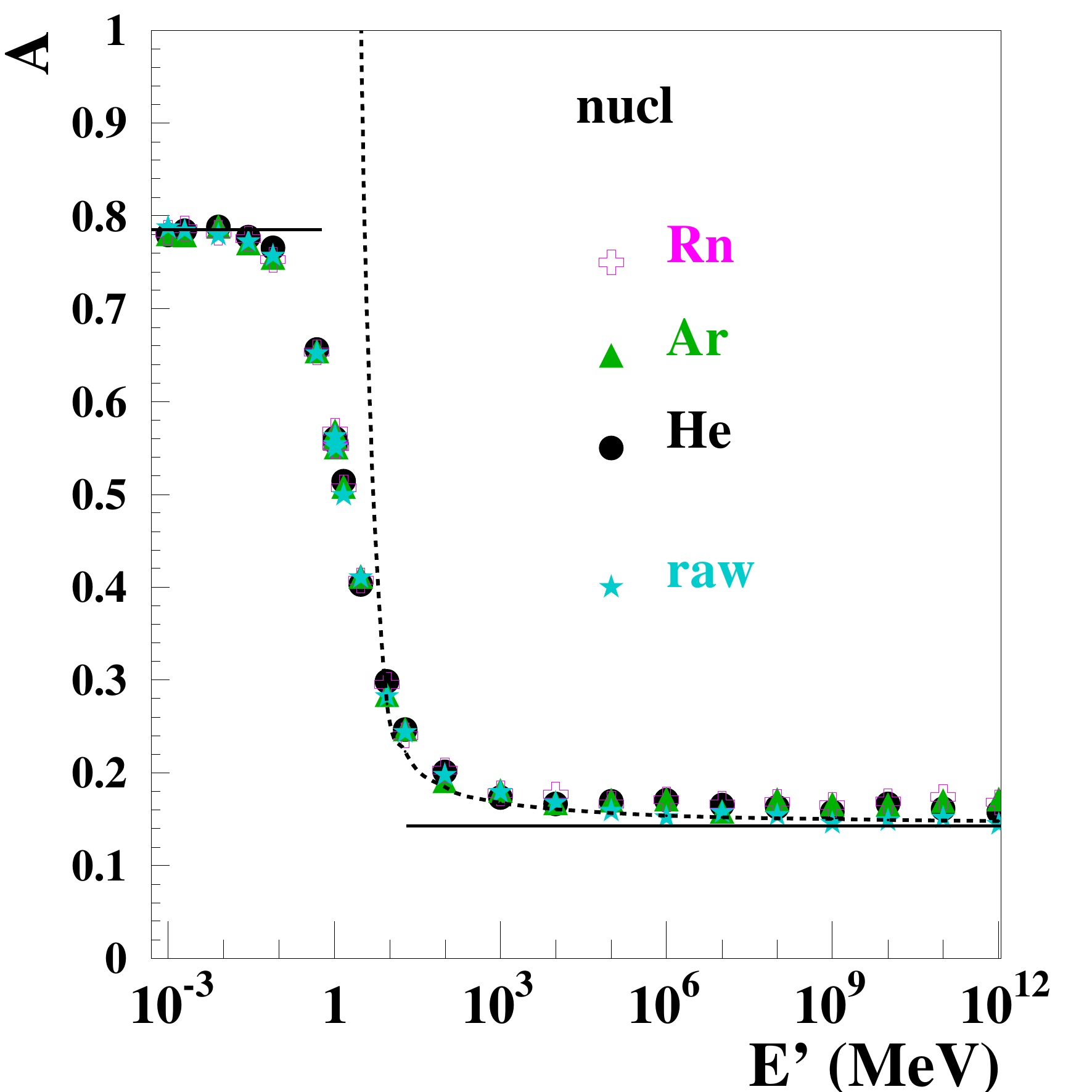}
 \includegraphics[width=0.49\linewidth]{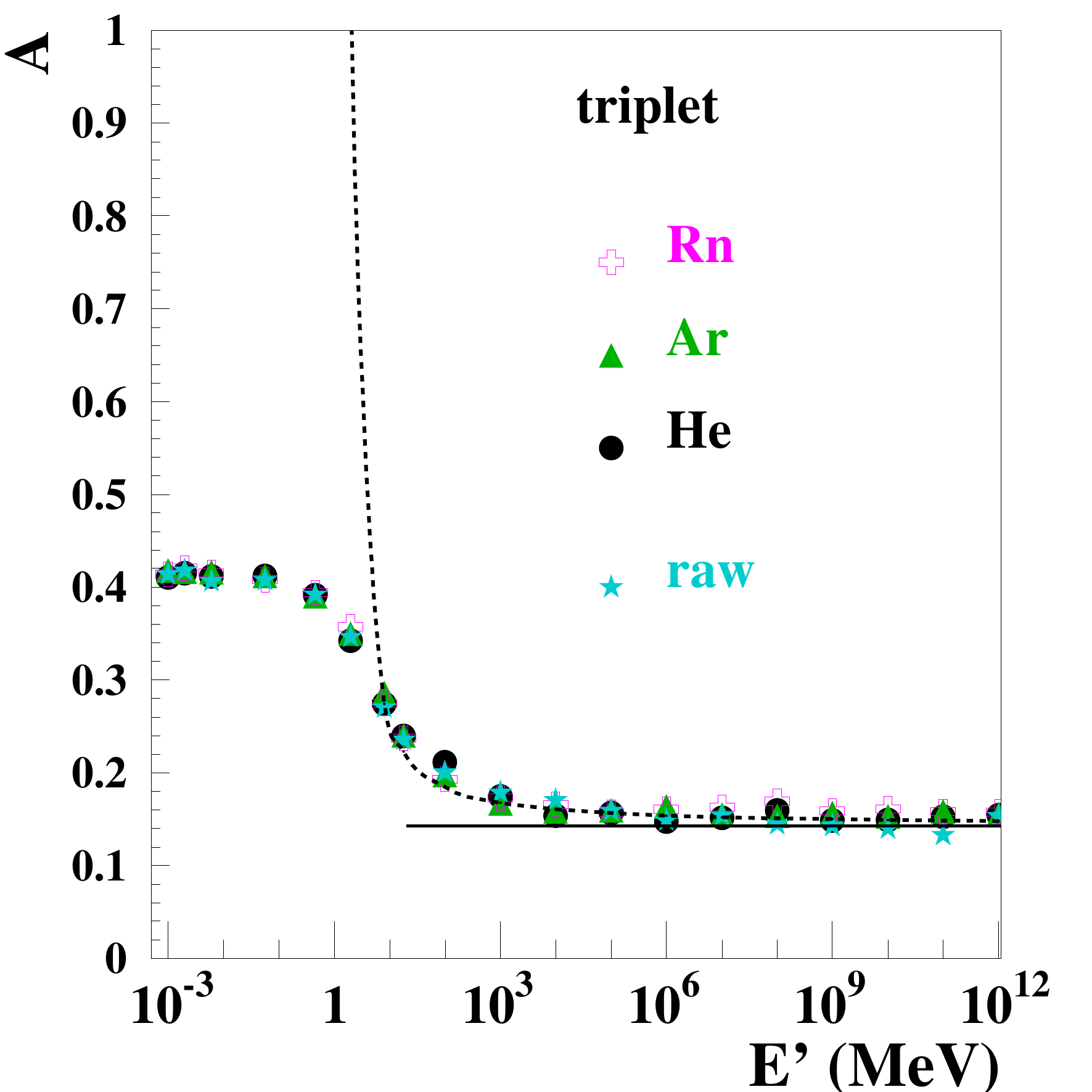}
\caption{Polarisation asymmetry calculated on event samples simulated
 with the new event generator, as a function of $E'$.
Left: nuclear conversion.
Right: triplet conversion.
``raw'' isolated charged target (star), and the following atoms:
helium (bullet),
argon (upper triangle),
radon (plusses).
The horizontal lines denote the low- and high-energy approximations
of $\pi/4$ and $1/7$, respectively.
The dashed curves denote the Boldyshev-Peresunko high-energy
approximation\,\cite{Boldyshev:1972va}.
\label{fig:asym}
 }
\end{figure}

\begin{figure}[ht] 
 \includegraphics[width=0.49\linewidth]{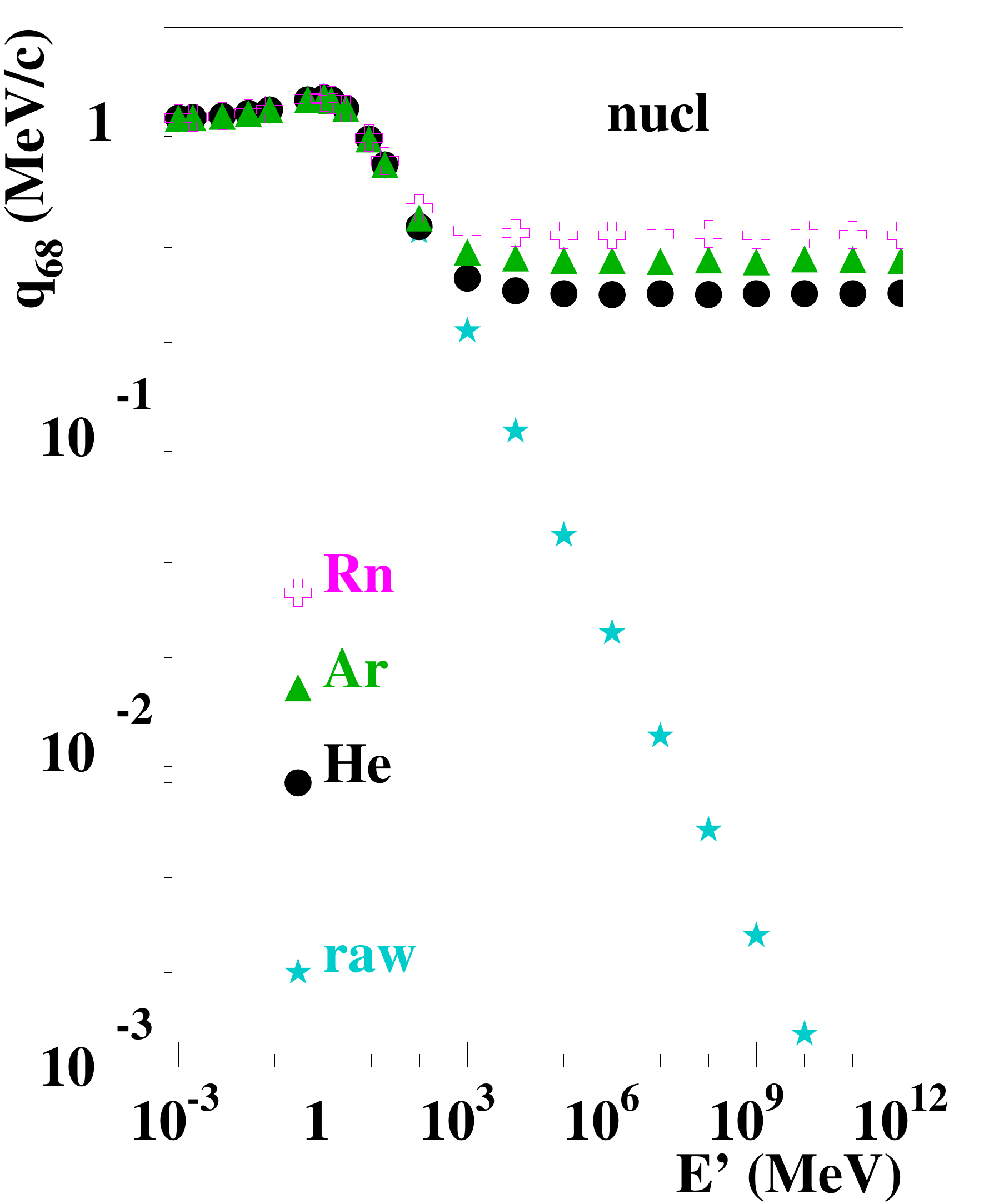}
 \includegraphics[width=0.49\linewidth]{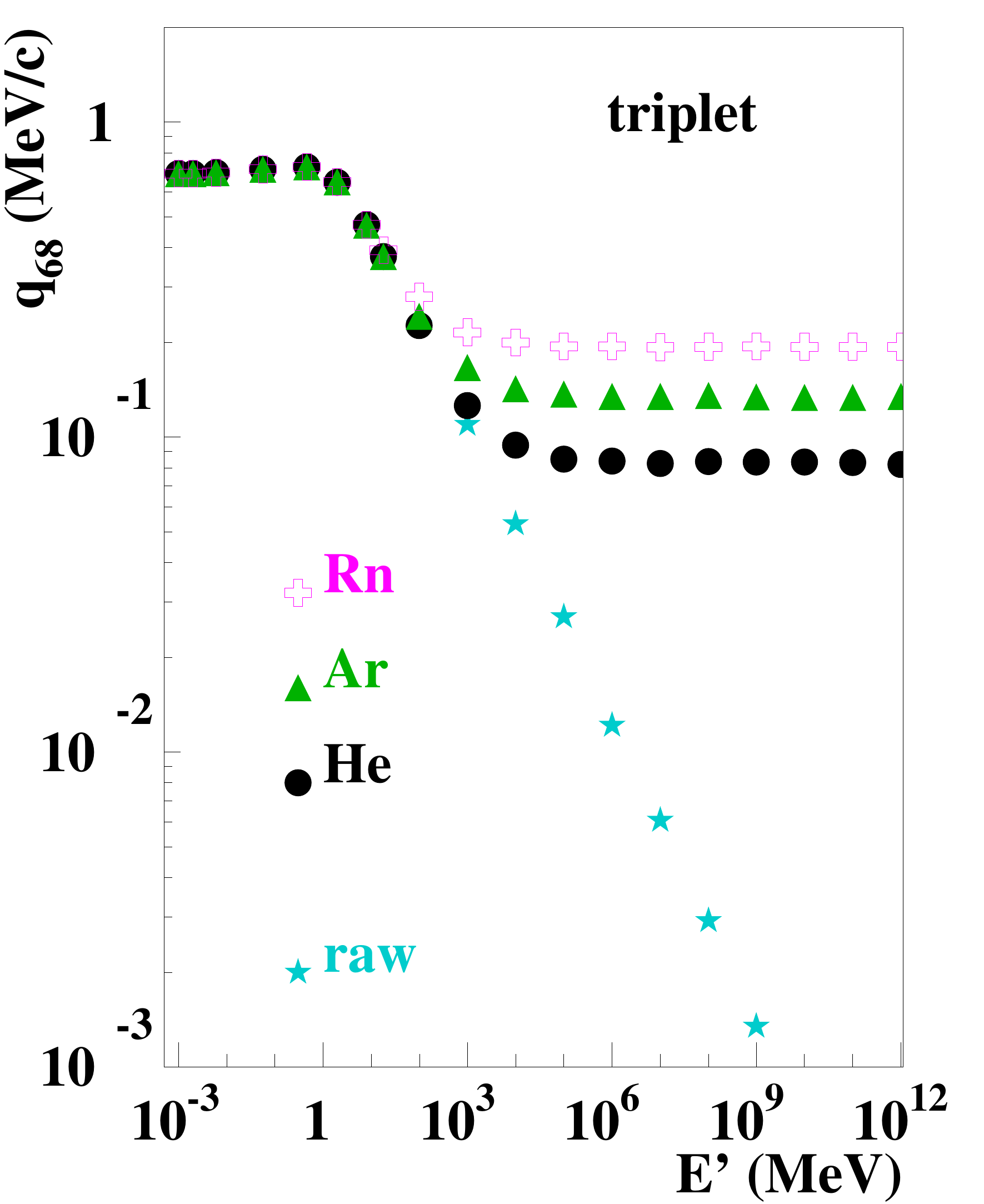}
 \caption{68\%-containment value $q_{68}$ of the recoil momentum $q$
 obtained from event samples simulated
 with the new event generator, as a function of $E'$.
Left: nuclear conversion.
Right: triplet conversion.
``raw'' isolated charged target (star), and the following atoms:
helium (bullet),
argon (upper triangle),
radon (plusses).
\label{fig:qfrac}
 }
\end{figure}

\begin{figure}[ht] 
 \includegraphics[width=0.49\linewidth]{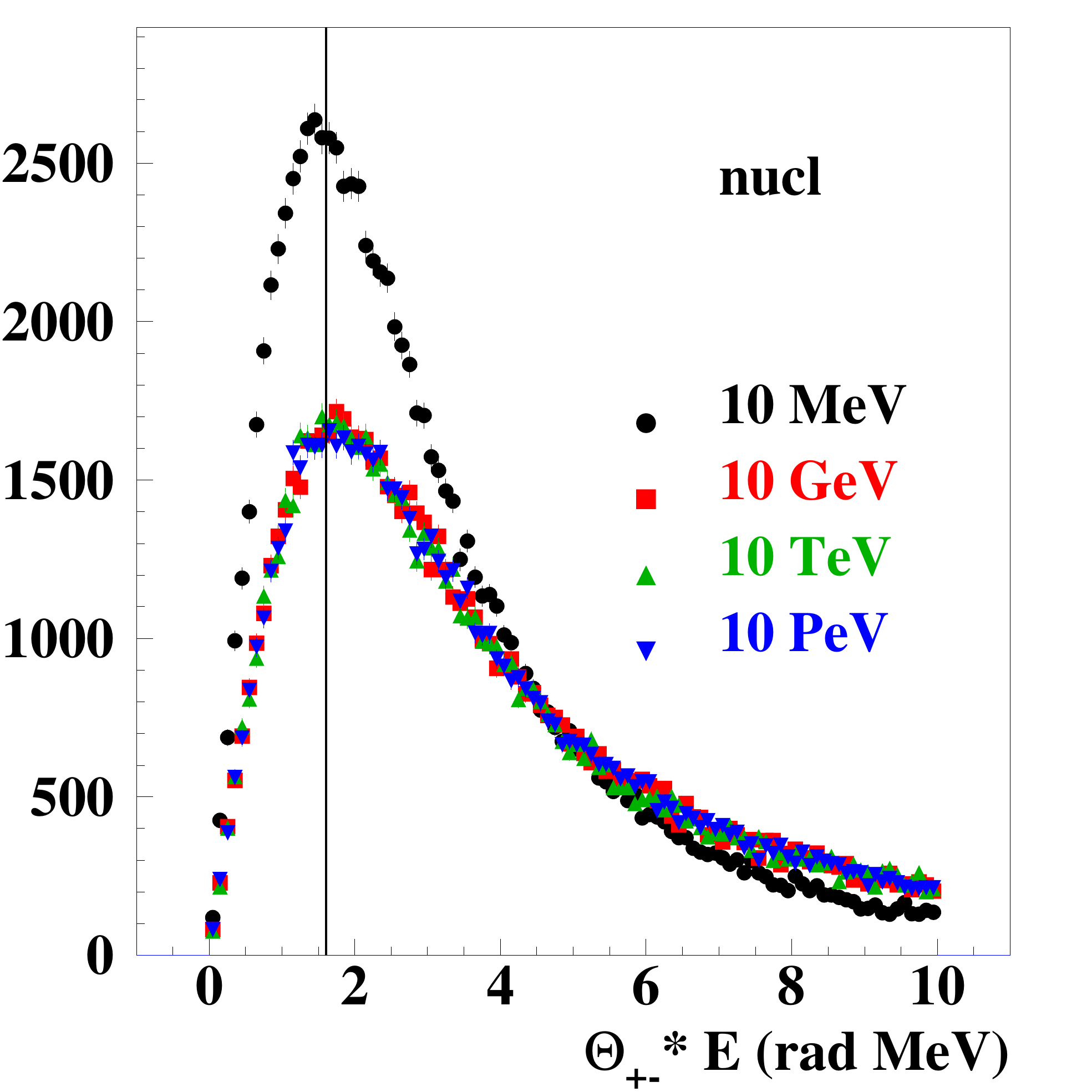}
 \includegraphics[width=0.49\linewidth]{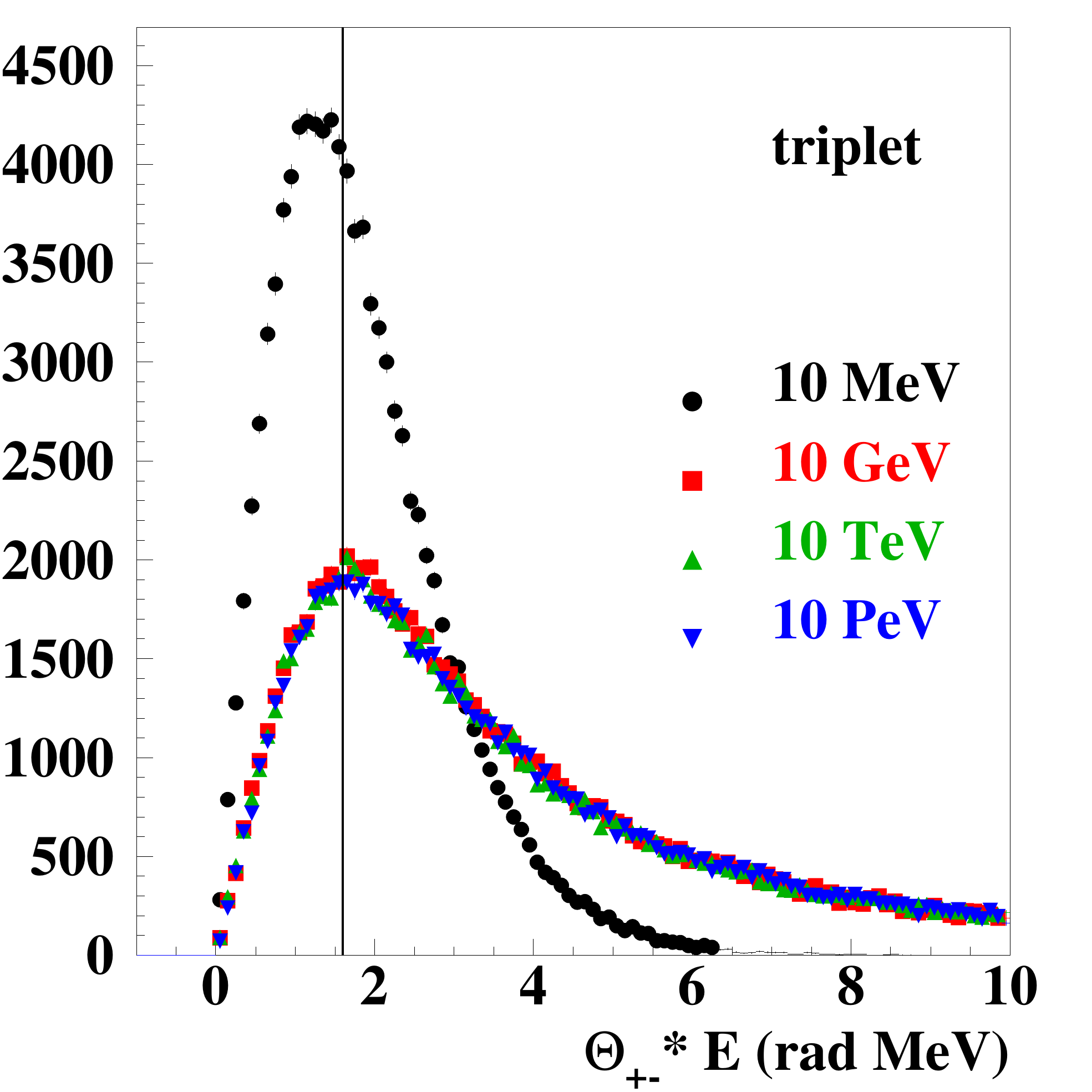}
 \caption{Distributions of the product of the pair opening angle and
 of the photon energy,
 $\theta_{+-} \times E$, for conversions on argon for
 10\,MeV (bullet),
 10\,GeV (square),
 10\,TeV (upper triangle) and
 10\,PeV (down triangle).
Left: nuclear conversion.
Right: triplet conversion.
The vertical value shows the most probable value of
$1.6\,\radian\,\mega\electronvolt$ computed by Olsen in the
high-energy approximation \cite{Olsen:1963zz}.
\label{fig:ouverture}
 }
\end{figure}

\begin{itemize}
\item The measurement of the polarisation angle $\varphi_0$ and of the
linear polarisation fraction $P$ of a gamma-ray beam 
 can be performed by the analysis of the distribution
of the event azimuthal angle $\varphi$ 
\begin{equation}
\gfrac{\dd N}{\dd \varphi} \propto 
\left(
1 + A \, P \cos[2(\varphi - \varphi_0)]
\right),
\label{eq:modulation}
\end{equation}
where $A$ is the polarisation asymmetry of gamma conversion to pairs.
Figure \ref{fig:asym} shows the polarisation asymmetry obtained from
the ($P=1$) samples, defining the event azimuthal angle as the
bisector of the electron and of the positron azimuthal angles,
$\varphi \equiv (\phi_++\phi_-)/2$ \cite{Gros:2016dmp}, and using the
moments' method \cite{Bernard:2013jea,Gros:2016zst,Gros:2016dmp}.
 
All (nuclear and triplet) ``raw'' results agree nicely with the
Boldyshev-Peresunko asympotic expression\,\cite{Boldyshev:1972va} at
high energy\footnote{We have corrected a misprint of
 \cite{Gros:2016dmp}, the normalisation of the photon energy to the
 electron rest mass energy in the log.}:
\begin{eqnarray}
A \approx \gfrac
{\gfrac{4}{9}\log{\left(2 E / m c^2\right)} - \gfrac{20}{28}}
{\gfrac{28}{9}\log{\left(2 E / m c^2\right)} - \gfrac{218}{27}}.
\label{eq:sig:HE}
\end{eqnarray}
At low energy the obtained values agree with the asympotic value
obtained in
\cite{Gros:2016dmp} for nuclear conversion but not for triplet
conversion.
\item As the distribution of the recoil momentum, $q$, has a long
 tail, (Fig. 3 of \cite{Bernard:2012uf}), the contribution to the
 photon angular resolution $\Delta \theta$ due to the fact that the
 nucleus recoil cannot be measured is not Gaussian distributed:
 astronomers make use of the 68\,\%-containment resolution angle.
 In the transverse-recoil approximation, which is valid for
$E' > 1\,\mega\electronvolt$ (Fig. 3 right of \cite{Gros:2016zst},
 $\Delta \theta$ is $\approx q/E$. Figure \ref{fig:qfrac} shows the
 68\%-containment value, $q_{68}$, of the recoil momentum, $q$,
 calculated on event samples simulated with the new event generator as
 a function of $E'$ and can be compared to Figs. 3, 5 and 6 of
 \cite{Gros:2016zst}.

\item Figure \ref{fig:ouverture} shows the distributions of the pair
 opening angle normalised to $1/E$ for conversions on argon.
 For $E > 10\,\mega\electronvolt$, they peak at the value of
$1.6\,\radian\,\mega\electronvolt$ computed by Olsen in the
high-energy approximation \cite{Olsen:1963zz}. 
\end{itemize}
 
\subsection{Applicability range}

The present results were obtained with REAL*16 machine precision, in
the energy range from 1\,keV above threshold up to 1\,EeV.
\begin{itemize}
 \item
For conversions on nuclei or on electrons bound in atoms, the generator
was found to provide nominal results, both with REAL*8 and REAL*16 machine
precisions.
 \item
For conversions on isolated charged particles and for REAL*8 machine
precision, the generator is found to compute the pdf wrongly
 below $q \approx 10^{-8}\,\mega\electronvolt/c$, which can be
reached for $\gamma$-ray conversions above
$E \approx 40\,\tera\electronvolt$.
Conversions on atoms are immune to this limitation because screening
prevents conversions with such low values of $q$.
\end{itemize}
Note that the CPU time increases with precision, from
$0.22\,\milli\second$/event (REAL*8) to 
$6.6\,\milli\second$/event (REAL*16)
for $100\,\mega\electronvolt$ $\gamma$-ray nuclear conversions
on argon,
on a DELL Precision M4600 machine.

\begin{figure}[ht] 
 \includegraphics[width=0.49\linewidth]{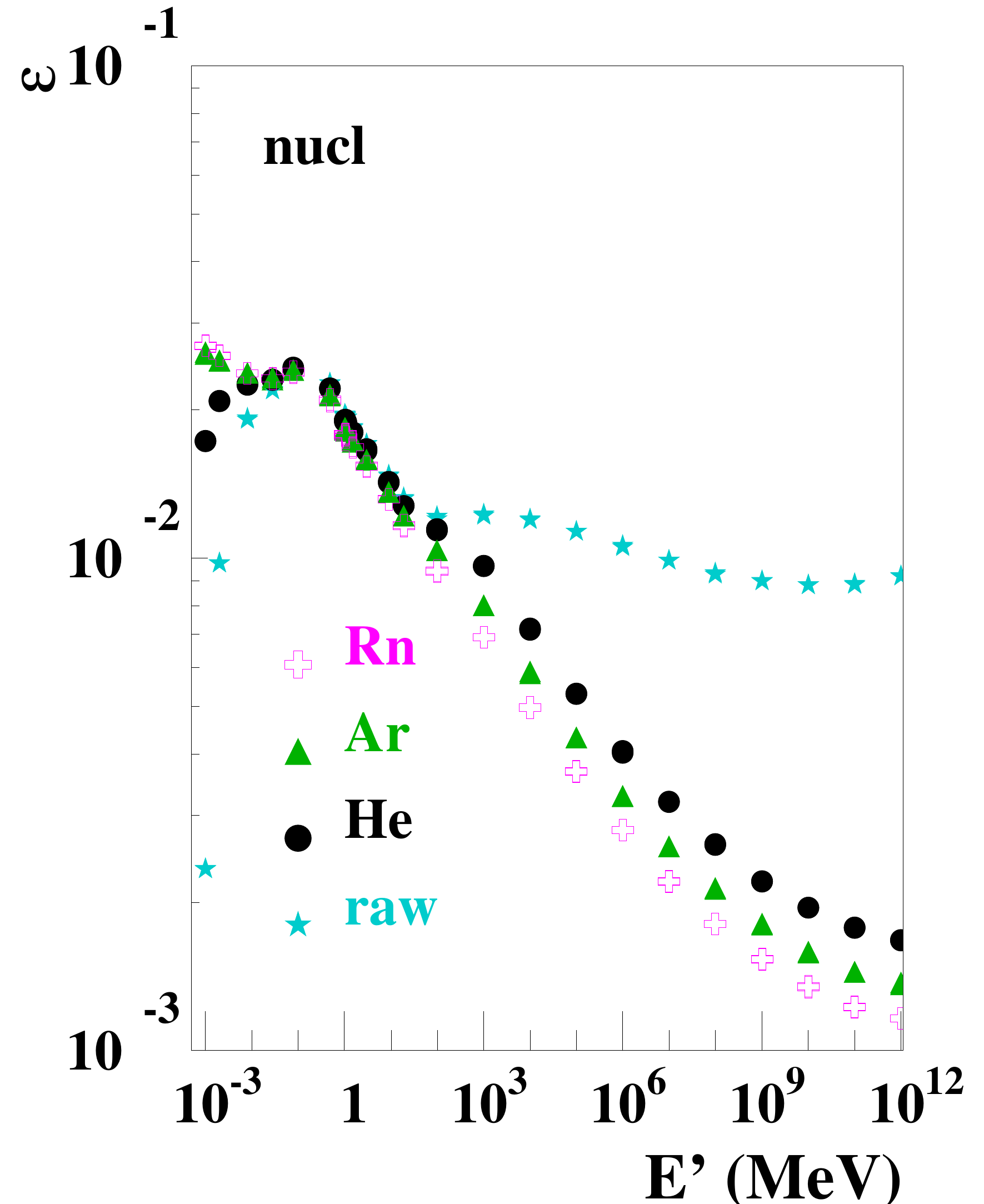}
 \includegraphics[width=0.49\linewidth]{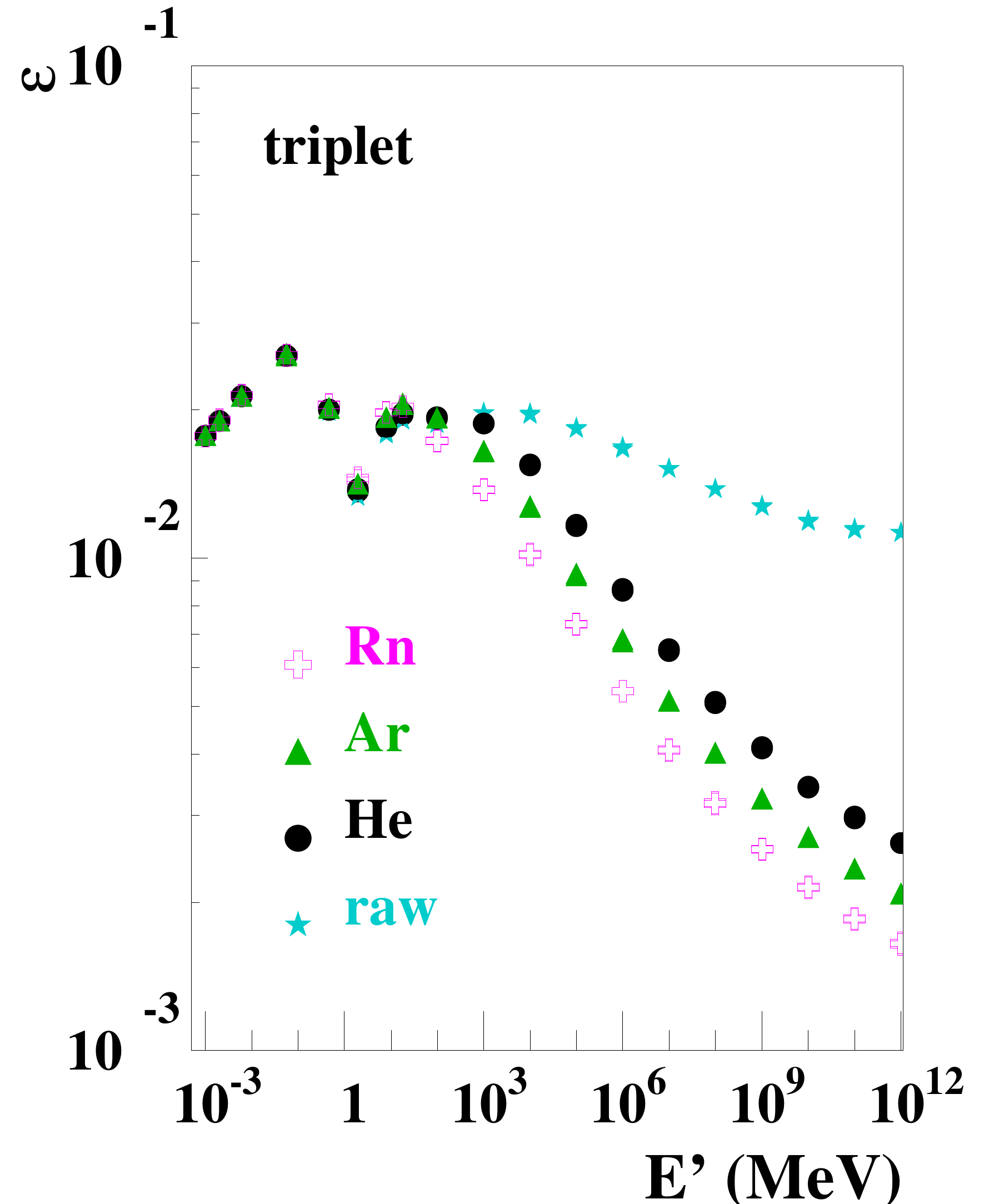}
 \caption{Efficiency of the new event generator, as a function of $E'$.
Left: nuclear conversion.
Right: triplet conversion.
``raw'' isolated charged target (star), and the following atoms:
helium (bullet),
argon (upper triangle),
radon (plusses).
\label{fig:jtry}
 }
\end{figure}

\subsection{Generator efficiency}

The efficiency of the generator, $\epsilon$, defined as the inverse of
the average number of computations of the pdf needed to obtain one
generated event, is low, between one-per-mille and several percent,
(Fig. \ref{fig:jtry}), which is a sign
of large correlations among the variables.
Therefore, it might be wise to restrict the use of this physical model
to the primary interaction of a photon in a detector and not to the
generation of a full EM shower.

We tried to find changes of variables on $x_1$ or $x_2$ to make the 1D
pdfs flatter in the hope of improving the overall generator
efficiency and we did not succeed.
So we took $x_1$ at random after a 1D pdf, which is equivalent to a
change of variable.

We first tried to take $x_1$ at random with pdf
$p_1(x_1) \equiv \int p(x) \dd x_2 \dd x_3 \dd x_4 \dd x_5 $,
where the integral is performed numerically.
We use the acceptance-rejection method described in
Sect. \ref{sec:common} on pdf $p(x) / p_1(x_1)$ instead of using it on
pdf $p(x)$, which is equivalent to a Jacobian factor correction.
We were not able to fine-tune this method because, due to the
already-mentioned
correlations, some events have a large value of $p(x)$
even though they have a small value of $p_1(x_1)$, so the value
of $p(x) / p_1(x_1)$ can be extremely high, making the determination
of $C$ inefficient.

Taking instead $x_1$ at random with pdf
$p_1'(x_1) \equiv \max_{x_2 \cdots x_5} p(x)$, a function that we
parametrised with parameters described as functions of $(E, Z)$
separately for nuclear and triplet conversions resulted in an
efficiency gain by a factor two to three in the $(E, Z)$ range
considered in this work.
This shows that indeed the low efficiency is intimately related to the
correlations amongst the variables.

\section{Conclusion}

We have developed a 5D, exact, polarised, Bethe-Heitler event
generator of $\gamma$-ray conversions to $e^+e^-$ that is able to simulate
successive events with different photon energies and different atomic
targets without any CPU overhead.
The strong correlation between kinematic variables in the divergence
of the five-dimensional differential cross section was mitigated by
performing each step of the conversion in the appropriate Lorentz
frame.
We have performed a number of verifications by comparison with
properties established in the past from analytical calculations, on
the photon energy range from 1\,keV above threshold up to 1\,EeV.
The calculation is currently implemented in fortran; it can be
implemented in other programming languages.

\section{Acknowledgments}

It is a pleasure to acknowledge the support of the French National
Research Agency (ANR-13-BS05-0002).

\end{document}